%% file: conference_101719.tex
\documentclass[conference]{IEEEtran}
\IEEEoverridecommandlockouts
\usepackage{cite}
\usepackage{amsmath,amssymb,amsfonts}
\usepackage{algorithmic}
\usepackage{subcaption}
\usepackage{graphicx}
\usepackage{textcomp}
\usepackage{xcolor}
\usepackage{comment}
\usepackage{hyperref}
\def\BibTeX{{\rm B\kern-.05em{\sc i\kern-.025em b}\kern-.08em
    T\kern-.1667em\lower.7ex\hbox{E}\kern-.125emX}}

\newcommand{\se}[1]{{\textcolor{black}{#1}}}
\newcommand{\secor}[2]{{\textcolor{black}{#2}}}
\newcommand{\secmt}[1]{}
    
\begin{document}

\title{Perceptual Noise-Masking with Music \\through Deep Spectral Envelope Shaping %*\\
%{\footnotesize \textsuperscript{*}Note: Sub-titles are not captured in Xplore and
%should not be used}
\thanks{This work was supported by the Audible project funded by French BPI, and was performed using HPC resources from GENCI-IDRIS (Grant 2023-AD011014883). 
}
}

% \author{\IEEEauthorblockN{Clémentine Berger}
% \IEEEauthorblockA{\textit{LTCI} \\
% \textit{Télécom Paris}\\
% Paris, France \\
% clementine.berger@telecom-paris.fr}
% \and
% \IEEEauthorblockN{Roland Badeau}
% \IEEEauthorblockA{\textit{LTCI} \\
% \textit{Télécom Paris}\\
% Paris, France \\
% roland.badeau@telecom-paris.fr}
% \and
% \IEEEauthorblockN{Slim Essid}
% \IEEEauthorblockA{\textit{LTCI} \\
% \textit{Télécom Paris}\\
% Paris, France \\
% slim.essid@telecom-paris.fr}
% }

\author{\IEEEauthorblockN{Clémentine Berger, Roland Badeau, Slim Essid}
\vspace{0.3cm}
\IEEEauthorblockA{
    \textit{LTCI}, \textit{Télécom Paris}, \textit{Institut Polytechnique de Paris}, Palaiseau, France \\
    \{firstname\}.\{surname\}@telecom-paris.fr
    }}

\maketitle

\begin{abstract}
%This document is a model and instructions for \LaTeX.
%This and the IEEEtran.cls file define the components of your paper [title, text, heads, etc.]. *CRITICAL: Do Not Use Symbols, Special Characters, Footnotes, 
%or Math in Paper Title or Abstract.
People often listen to music in noisy environments, seeking to isolate themselves from ambient sounds. Indeed, a music signal can mask some of the noise's frequency components due to the effect of simultaneous masking. In this article, we propose a neural network based on a psychoacoustic masking model, designed to enhance the music's ability to mask ambient noise by reshaping its spectral envelope with predicted filter frequency responses. The model is trained with a perceptual loss function that balances two constraints: effectively masking the noise while preserving the original music mix and the user's chosen listening level. We evaluate our approach on simulated data replicating a user's experience of listening to music with headphones in a noisy environment. The results, based on defined objective metrics, demonstrate that our system improves the state of the art.
\end{abstract}


\section{Introduction}
\input{sections/intro}

\section{Neural model}
\input{sections/neural_model}

\section{Data}
\input{sections/datas}

\section{Results}
\input{sections/results}

\section{Conclusions}
\input{sections/conclusions}

%\section*{Acknowledgment}

%The preferred spelling of the word ``acknowledgment'' in America is without 
%an ``e'' after the ``g''. Avoid the stilted expression ``one of us (R. B. 
%G.) thanks $\ldots$''. Instead, try ``R. B. G. thanks$\ldots$''. Put sponsor 
%acknowledgments in the unnumbered footnote on the first page.

\bibliographystyle{IEEEtran}
%\bibliography{IEEEabrv,bibliography}    
\bibliography{biblio}    

\end{document}

%% file: sections/intro.tex
Listening to music %with headphones or earphones 
in noisy environments may negatively impact the listening experience as the surrounding noise interferes with the perception of the music, partially masking some spectral components of the audio content \cite{mooreModelPredictionThresholds1997, ramoPerceptualFrequencyResponse2012}. A part of the music may even be completely concealed by the noise due to simultaneous masking. For a given audio signal, this effect is quantified using masking thresholds, which indicate the level below which another signal becomes inaudible within specific frequency bands due to the presence of the first signal \cite{painterPerceptualCodingDigital2000, zwickerPsychoacousticsFactsModels2010}.

Thus, music rendering systems have been developed over the years to enhance listening comfort, initially in automotive environments \cite{clarkCompensationRoadNoise1987,millerCopingRoadNoise1994,christophNoiseDependentEqualization2012d} and later for more general contexts and personal devices such as headphones or earphones \cite{ramoPerceptualHeadphoneEqualization2013, jrAdaptedAudioMasking2015,jrCollaborativelyProcessingAudio2016}.  One approach involves volume adjustments and compression to increase the loudness of the music when it is masked or partially masked by noise \cite{clarkCompensationRoadNoise1987, millerCopingRoadNoise1994, jrAdaptedAudioMasking2015, kuivalainenAdaptiveModulationAudio2021}.  Similarly, adaptive perceptual equalizers have been proposed to restore the original loudness of the music signal when it is affected by ambient noise \cite{christophNoiseDependentEqualization2012d, ramoPerceptualHeadphoneEqualization2013}.

However, the masking effect works both ways: music may also be used to mask ambient noise. Some works on Active Noise Cancelling (ANC) systems have utilized psychoacoustic information to focus on the frequency bands where noise exceeds the music's masking thresholds, ensuring that the residual noise after reduction is completely masked by the music \cite{docloActiveNoiseReduction2016, belyiIntegratedPsychoacousticActive2019, zachosFeedforwardHeadphoneActive2024}. Another strategy is to filter the music so that its masking thresholds are raised above the noise level. The main challenge is then how to generate filtering parameters that achieve these perceptual goals while maintaining the original music's identity and the user’s preferred listening level. Estreder et al. \cite{estrederPerceptualAudioEqualization2018} proposed a system that computes the gain parameters of a graphic equalizer whose goal is to boost the frequency bands where the noise is not masked. However, they simplified the problem by not fully leveraging the psychoacoustic model in the gain computation and not incorporating constraints to minimize music alteration, aside from limiting the gain in each band.

In this article, we propose an original deep neural network approach to produce filters that enhance the played-back music's capacity to mask ambient noise. Like Estreder et al. \cite{estrederPerceptualAudioEqualization2018}, we focus on simultaneous masking, excluding the partial masking effect.  We demonstrate that using a well-chosen neural network with a well-designed loss function allows a better exploitation of the psychoacoustic model for gain prediction and the integration of other criteria to balance masking effectiveness and musical fidelity. \se{Specifically, our contributions are: i) a U-net architecture optimized with a perceptual loss function trading-off masking and minimal power variations leading to improved results over the method proposed by Estreder \cite{estrederPerceptualAudioEqualization2018}; ii) a perceptually motivated procedure to deduce equalization frequency responses from predicted gains in each frequency band; iii) an experimental design studying variants of the proposed system based on different levels of power constraint, using data simulated with headphones responses to reproduce realistic auditory scenes, and evaluation metrics jointly assessing masking and power-preservation.} This study focuses on the context of music listened to with headphones; however, the proposed method is more general and can be applied to other use cases.

%% file: sections/neural_model.tex
\begin{figure}
    \centering
    \includegraphics[width=\linewidth]{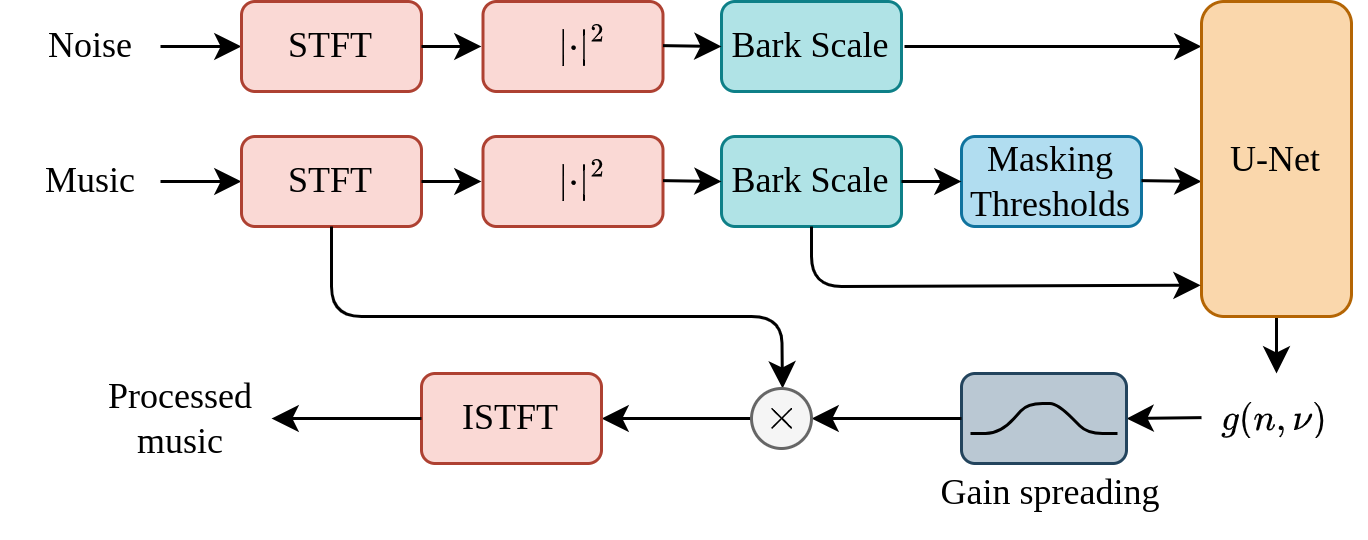}
    \caption{Overview of the proposed system. Bark features from the music and noise signals are computed: PSD per Bark band for both music and noise and music masking thresholds. The features are fed to the U-Net that outputs gains in dB used to scale filter frequency responses. They are applied in the spectral domain to the music and a processed version is generated by inverse STFT.}\vspace{-0.3cm}
    \label{fig:system_overview}
\end{figure}

In this work, we tackle the task of raising the listened music's masking thresholds to better mask the listener's surrounding noise.
The proposed model Deep Perceptual Noise Masking with Music (DPNMM) is based on a deep neural network that predicts the frequency responses of filters to apply to the music to achieve the desired masking effect. A schematic overview of the system is shown in Fig. \ref{fig:system_overview}. The neural network takes as input frequency features derived from both the noise and the music. Their respective power spectral densities (PSDs) are computed using sliding Hanning windows of length $N_{win}=$ 2048 points with a 75\% overlap and a sampling rate $f_s=$ 44100 Hz.  
Both are mapped to the psychoacoustic Bark scale \cite{zwickerPsychoacousticsFactsModels2010} resulting in the PSDs $P^{noise}(n, \nu)$ and $P^{music}(n, \nu)$, where $n$ denotes the frame index and $\nu$ the Bark band. 

The music masking thresholds $T(n, \nu)$ per Bark band are computed using Johnston's model \cite{johnstonTransformCodingAudio1988} as in Estreder's system \cite{estrederPerceptualAudioEqualization2018}. The calculation involves applying a \textit{spreading function} to the PSD in each critical band, which indicates how it masks signals within the same band and in adjacent bands. The contributions from all Bark bands are then summed, following the additivity property of simultaneous masking \cite{lutfiAdditivitySimultaneousMasking1983, johnstonTransformCodingAudio1988, humesModelsAdditivityMasking1989}, and adjusted downward by an offset that depends on whether the music is more tonal or noise-like.

\subsection{Architecture}

\begin{figure}
    \centering
    \includegraphics[width=0.9\linewidth]{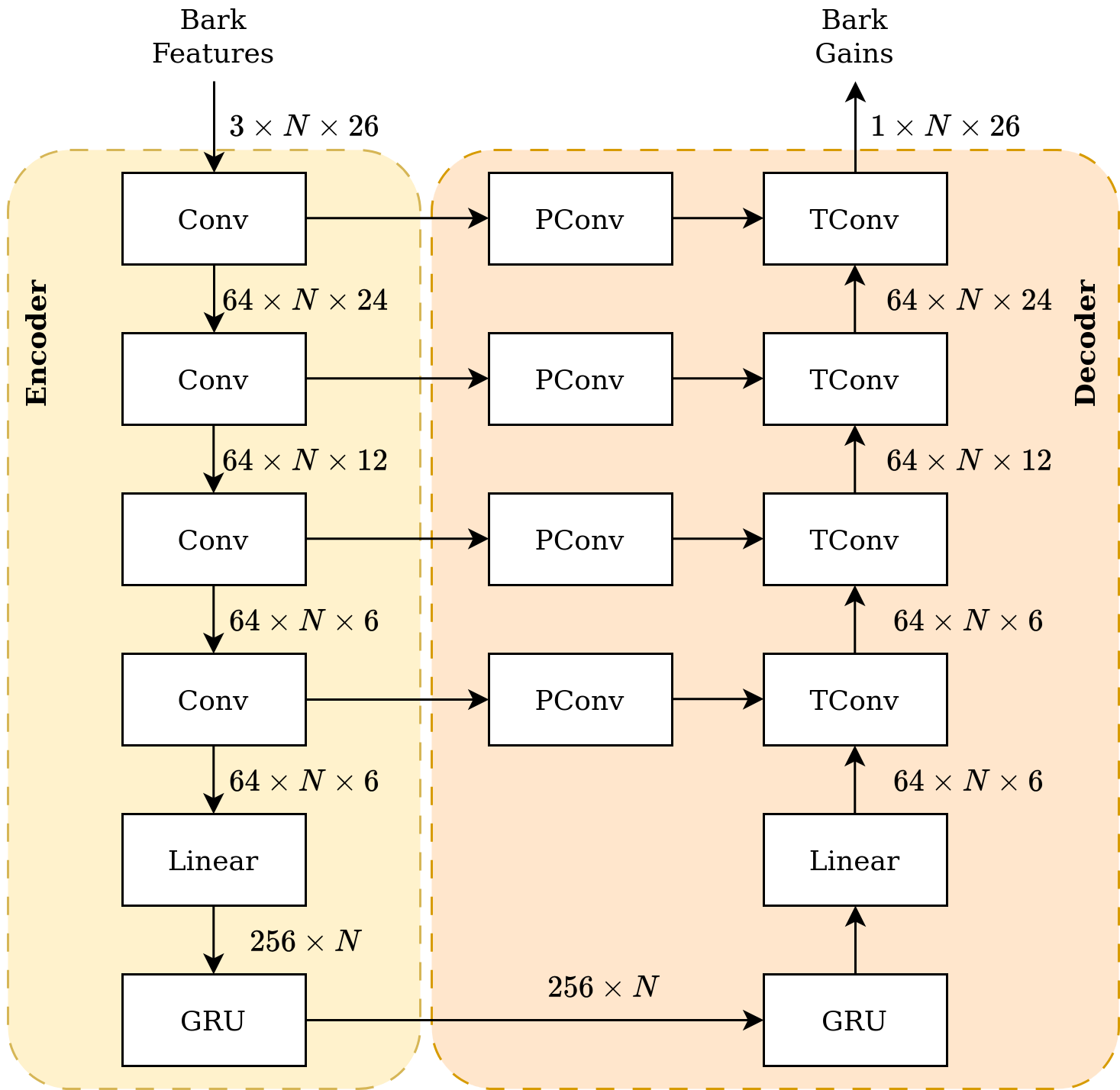}
    \caption{U-Net architecture of the proposed model. $N$ is the number of time frames, and 26 the number of Bark bands. The encoder is composed of 4 convolutional layers (Conv), a linear layer and a Gated Recurrent Unit (GRU) layer. The decoder follows the inverse path with transposed convolutional layers (TConv) and pathway convolutions (PConv) as add-skip connections.}%\vspace{-0.3cm}
    \label{fig:u-net_architecture}
\end{figure}

Our model's architecture is illustrated in Fig. \ref{fig:u-net_architecture}. We design a U-Net encoder-decoder model, inspired by the DeepFilterNet architecture \cite{schroterDeepfilternetLowComplexity2022}, especially its Equivalent Rectangular Bandwidth (ERB) gains prediction branch. The encoder consists of 4 convolutional blocks (separable convolution + BatchNorm + ReLU) followed by a linear layer, designed to process frequency information between the Bark features. 
%while preserving the temporal structure. 
This frequency information is then passed through a Gated Recurrent Unit (GRU) layer to handle the temporal dynamics. The decoder mirrors the encoder and incorporates skip connections, outputting the predicted gains in dB, $g(n,\nu)$, per frame, and Bark band (up to the 24th band $\sim$ 16500 Hz). 
Additionally, a gain smoothing filter is applied to prevent too rapid variations in the filters' frequency responses over time, and the gains are finally clamped as in \cite{estrederPerceptualAudioEqualization2018} with the following threshold values:
\begin{equation}
    g(n, \nu) \leftarrow \max \big( \min( g(n, \nu); -5 \text{ dB} ) ; 10 \text{ dB} \big).
\end{equation} 
\vspace{-0.5cm}

\subsection{Filters frequency responses}

\begin{figure}
    \centering
    \includegraphics[width=\linewidth]{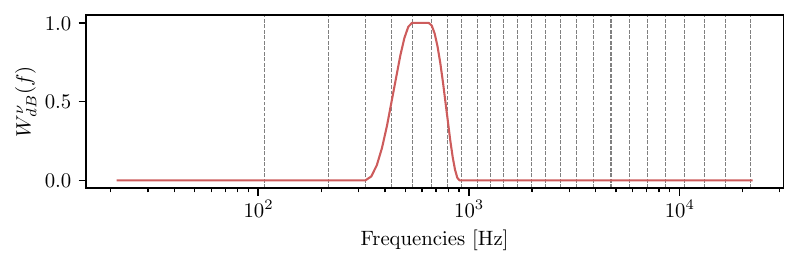}
    \caption{Pattern $W_{dB}^\nu(f)$ used to shape the filters' frequency responses, for $\nu =$ 5.}\vspace{-0.3cm}
    \label{fig:pattern_filter}
\end{figure}

The obtained gains are converted into filter frequency responses
that are generated using a pattern $W_{dB}^\nu(f)$ centered on the corresponding Bark band $\nu$ (see Fig. \ref{fig:pattern_filter}). In each band, the pattern is constant and equal to 1. 
To simulate the spreading effect of masking across the Bark bands, the pattern transitions smoothly with a cosine shape across the two adjacent bands below (from 0 to 1) and above (from 1 to 0). For the lowest frequency band, the pattern level is set to 2 instead of 1 at the center. This approach accounts for the fact that adjacent bands also contribute to raising the masking threshold of a given band, thereby distributing the required gain across multiple bands and minimizing the boost needed on the central band. %alone.

For each audio frame the overall frequency response $\mathcal{W}_{dB}(n,k)$ is obtained as:
\begin{equation}
    \mathcal{W}_{dB}(n,k) = \sum_{\nu=1}^{B} g(n,\nu) \cdot W_{dB}^\nu(k).
\end{equation}
The input music is then filtered, in the frequency domain, using those frequency responses frame by frame to produce a processed music whose masking properties are enhanced.

\subsection{Loss functions}
Our model DPNMM is trained with a primary loss function designed to raise the music's masking threshold above the noise level in the critical bands where needed. These Bark bands are constrained to exceed specific thresholds, while the energy in other bands is free to change as long as their masking threshold remains above the noise level, thereby supporting masking in adjacent bands: 
\begin{equation}
    \mathcal{L}_0 (\theta_t) = \frac{1}{N} \frac{1}{B}  \sum_{n,\nu} \text{ReLU}\left( P_{dB}^{noise}(n,\nu) - \hat{T}_{dB}(n,\nu) \right) ,
\end{equation}
where $\hat{T}_{dB}(n,\nu)$ is the masking threshold computed with the processed music, $N$ the number of time frames and $B$ the number of Bark bands.
By using a ReLU function to express the constraint, we only set a minimum threshold level for the network to reach, allowing it greater flexibility to find solutions for masking noise across all frequency bands. However, while this choice aims to provide the network with more freedom, it also brings the challenge that the system is not required to output zero gains when no amplification of the music is needed. To address this, we use the knowledge of the masking spreading effect  to compute a mask that identifies, for each band, whether it is close enough to another band (including itself) where the threshold needs to be raised to have an impact on it. If it is not, the gain for that band at the network's output is set to zero.

Even with this gain masking, the system still has an infinite number of potential solutions. To guide the learning process in a desired direction, we add a secondary constraint aimed at preserving the naturalness of the original music by limiting the average power variation: 
\begin{equation}
    \mathcal{L}_{power}(\theta_t) = \frac{1}{N} \sum_n \vert \hat{\mathcal{P}}_{dBA}^{music}(n) - \mathcal{P}_{dBA}^{music}(n) \vert \; ; 
\end{equation}
where the initial music mean power $\mathcal{P}$ of frame $n$ is evaluated in dBA \cite{a_weights}, as well as the processed music mean power $\hat{\mathcal{P}}$. To include this constraint in the training process we use a strategy inspired by the method of multipliers \cite{jonasdegraveHowWeCan2021, plattConstrainedDifferentialOptimization1987}. The goal is to ensure that during training the power variation does not exceed a given value $\Delta \mathcal{P}_{max}$ using a dynamic weight $\lambda_t$ to scale the loss. The overall loss function for this constrained optimization problem is:
\begin{equation}
    L(\theta_t, \lambda_t) = \mathcal{L}_0 - \lambda_t \cdot (\Delta \mathcal{P}_{max} - \mathcal{L}_{power}(\theta_t)).
\end{equation}
At the start of the training, $\lambda_t = 0$. While gradient descent is applied to the overall loss $L(\lambda_t)$, gradient ascent is simultaneously performed on $\lambda_t$ using the gradient $\frac{\partial L(\theta, \lambda_t)}{\partial \lambda_t} = \mathcal{L}_{power}(\theta_t) - \Delta \mathcal{P}_{max}$ with a specific learning rate of $10^{-3}$, keeping $\lambda_t$ always positive. When $\mathcal{L}_{power}(\theta_t)$ exceeds the threshold $\Delta \mathcal{P}_{max}$, $\lambda_t$ increases, giving more weight to the power constraints in the total loss. Conversely, when $\mathcal{L}_{power}(\theta_t)$ drops below $\Delta \mathcal{P}_{max}$, $\lambda_t$ decreases.
This approach allows us to guide the training in a direction that satisfies both constraints without requiring a tedious search for an optimal fixed weight. The degree of compromise between the two constraints is directly controlled by the choice of parameter $\Delta \mathcal{P}_{max}$. In the rest of the article, we explore several values for this parameter. 

The code for the implementation of DPNMM is available on github.\footnote{\href{https://github.com/ClementineBerger/DPNMM}{https://github.com/ClementineBerger/DPNMM}}

%% file: sections/datas.tex
To train and evaluate the proposed model, we generated training, validation, and test datasets that replicate realistic acoustic scenes. These scenes \secor{consist of music listened to by a user through headphones or earphones and ambient noise as perceived by the user through their earphones}{simulate a user listening to music through headphones or earphones while being in a noisy environment}.  

We defined several \textit{environments} to represent a variety of realistic acoustic scenes with different ambient noise levels: urban, indoor office, construction site, beach, transportation (train/plane/boat), and restaurant/bar. 
\se{We chose} the noise recordings from the DNS Challenge dataset \cite{dubeyIcassp2022Deep2022}.
Each environment is \se{created using} the labels from the noise dataset and a realistic noise level distribution in dBA. Noise samples are evenly selected per environment and normalized to levels sampled from the corresponding noise distribution. A pre-processing step is applied to drop the audio signals composed mainly of silence (which are therefore isolated, impulsive noises). After that, all those samples are filtered with one of three headphone frequency responses to reproduce their passive attenuation. 
Each noise sample is then paired with a music track \se{chosen from the FMA dataset \cite{defferrardFMADATASETMUSIC2017}} which covers a large diversity of music genres (Pop, Rock, Classical, Jazz, Hip-Hop, etc.).
Each music is normalized to a dBA level derived from a Signal-to-Noise Ratio (SNR) value, itself sampled from a defined SNR distribution. The resulting music dBA level is constrained within the \secor{45 dBA to 100 dBA range}{range $[45, 100]$ dBA,} reflecting the typical range offered by standard headphones.
\se{We thus generate} 50h of training data, 20h of validation data, and 10h of test data, composed of pairs of 10s mono music and noise excerpts sampled at 44100 Hz.

%% file: sections/results.tex
\subsection{Baseline and evaluation metrics}

We use as a baseline the perceptual equalizer used by Estreder et al. \cite{estrederPerceptualAudioEqualization2018} that is based on a Bark band graphic equalizer employing second-order peak filters \cite{abelFilterDesignUsing2004a, valimakiAllAudioEqualization2016}.
The gains per Bark band are computed in order to raise the music level in the bands where the noise PSD is above its masking threshold,
\begin{equation}
    g(n,\nu) = \max \left( P_{dB}^{noise}(n,\nu) - T_{dB}(n,\nu); 0 \right) .
    \label{eq:estreder}
\end{equation}

This computation approach does not consider the additivity property of simultaneous masking.  The level in the $\nu$-th Bark band is raised to bring the masking threshold at the noise level by adjusting only this band while raising the levels in adjacent bands would also contribute to increasing the masking threshold in the target band via the effect of the spreading function.

Two objective metrics are considered to perform the evaluation of the models. We compute a mean Noise-to-Mask Ratio (NMR) per audio sample of the test set, only selecting the Bark bands where the initial music masking threshold is below the noise level :
\begin{equation}
\text{NMR} = \frac{1}{M} \sum_{n, \nu} (1-m_\nu(n)) | P_{dB}^{noise}(n,\nu) - \hat{T}_{dB}(n,\nu) |,
\end{equation} 
where $M = \sum_{n, \nu} (1-m_\nu(n))$ with $m_\nu$ a mask such that $m_\nu(n) = 0$ if the initial threshold is below the noise, and $m_\nu(n) = 1$ otherwise. The obtained NMR is compared to the initial NMR with the unprocessed music to evaluate how much the system can improve the masking effect on the bands where it is required. However, the system may as well induce power variations in the other bands. To evaluate this effect we also compute a mean Global Level Difference (GLD) :

\begin{equation}
    \text{GLD} = \frac{1}{N} \sum | \hat{\mathcal{P}}_{dBA}^{music}(n) - \mathcal{P}_{dBA}^{music}(n) | .
\end{equation}
Both metrics are computed by frequency \secor{bands}{ranges}: broadband, first \secor{tier}{third} of Bark bands (low), second third (medium), and last third (high). 

\subsection{\se{Discussion}}
\begin{figure}
\centering
\begin{subfigure}{\linewidth}
    \includegraphics[width=\textwidth]{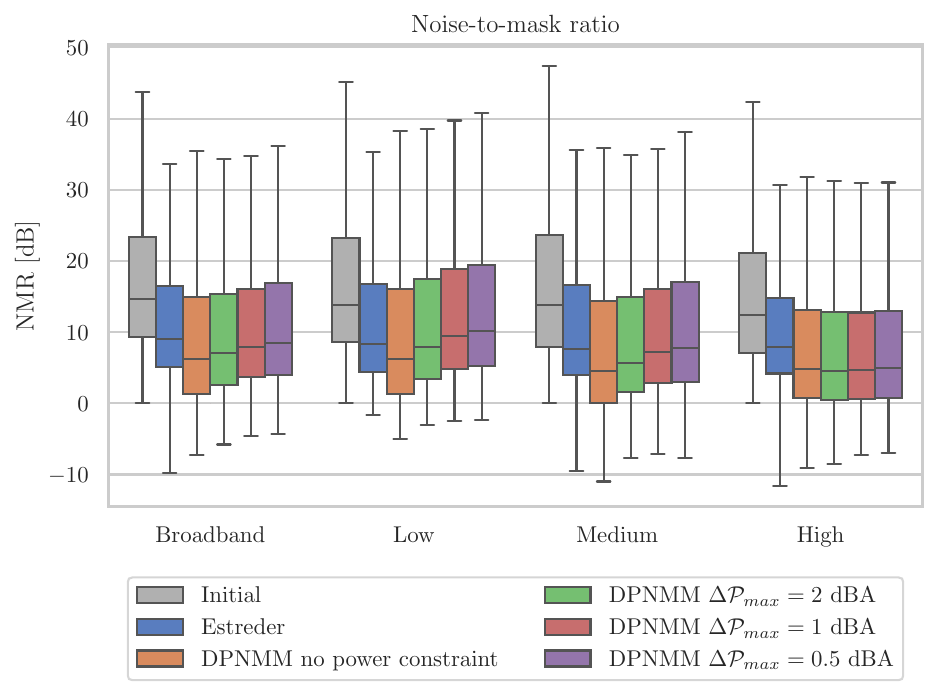}
    \label{fig:nmr}
\end{subfigure}
%\hfill
\begin{subfigure}{\linewidth}
    \includegraphics[width=0.98\textwidth]{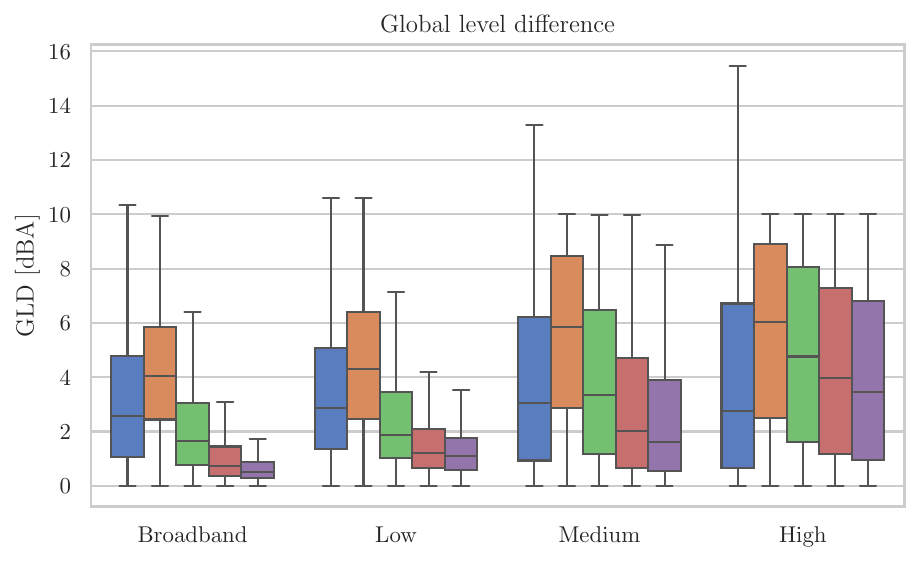}
    \label{fig:gld}
\end{subfigure}
        
\caption{Obtained NMR and GLD on the test set for the Estreder model and four versions of the proposed neural model with different degrees of power constraint during training : no constraint and $\Delta \mathcal{P}_{max} = 2, 1, 0.5$ dBA.}\vspace{-0.3cm}
\label{fig:results}
\end{figure}

We evaluate our approach by testing different configurations of the power constraint: no power constraint, $\Delta \mathcal{P}_{max} =$ 2, 1, and 0.5 dB. In all configurations, the model is trained for 50 epochs with a learning rate of $10^{-3}$ and a batch size of 64. The statistical difference between each of DPNMM configurations and the baseline model is evaluated by calculating the mean \textit{p}-value with the Wilcoxon test over 100 batches of 50 samples from the test set, with a Bonferroni correction applied.
The results are presented in Fig. ~\ref{fig:results}. 

In terms of NMR, all three versions of the neural model outperform Estreder's model on the broadband metric statistically significantly, except DPNMM with $\Delta \mathcal{P}_{max} = 0.5$ dBA. The version of the neural model without any power constraint performs the best compared to the baseline (\textit{p}-value $= 7\cdot 10^{-8}$). Applying a power constraint results in a decrease in performance all the more important the stricter the constraint (low $\Delta \mathcal{P}_{max}$), particularly in the low-frequency range to the point of becoming less performant than Estreder's PEQ. This outcome is expected, given the relatively low weight of high frequencies in the power measurement. When the power constraint is strict, the low and mid frequencies are more significantly affected. This trend is confirmed when examining the GLD. Without a power constraint, the neural model achieves excellent NMR performance by significantly amplifying the musical signal compared to Estreder's model. Adding the power constraint has then a clear beneficial effect on the GLD measure, \secor{even allowing}{thus achieving significantly} better results compared to the baseline model\secmt{you need to do statistical testing and provide p-value; use Wicoxon signed rank, see mattermost}, except at high frequencies where the model is less affected by the constraint. In particular, both neural models with constraints $\Delta \mathcal{P}_{max} =$ 2, 1 dBA achieve a better NMR than Estreder's model (\textit{p}-value of $10^{-6}$ and 0.01) and a better broadband GLD (\textit{p}-value of $1.5 \cdot 10^{-4}$ and $3.3 \cdot 10^{-9}$).

Audio examples and further results are available on this article's companion website.\footnote{\href{https://clementineberger.github.io/DPNMM/}{https://clementineberger.github.io/DPNMM/}}

%% file: sections/conclusions.tex
We proposed a new model, DPNMM, for estimating filters to shape music and raise its masking thresholds according to the spectral characteristics of ambient noise, in the context of users wearing headphones. This approach is perceptually motivated and allows balancing the effectiveness of the masking effect with a power-preservation constraint to ensure fidelity to the original recording.
This model shows superior performance compared to the baseline PEQ by Estreder et al. \cite{estrederPerceptualAudioEqualization2018} outperforming it both in terms of noise masking and power-preservation. 
Future works will explore incorporating user preferences to adapt the model's response to the user's preferred sound rendering.